# INTEGRATION OF KNOWLEDGE TO SUPPORT AUTOMATIC OBJECT RECONSTRUCTION FROM IMAGES AND 3D DATA

*Frank Boochs[1], Andreas Marbs[1], Helmi Ben Hmida[1,2], Hung Truong[1,2], Ashish Karmacharya[1,2], Christophe Cruz[2], Adlane Habed[2], Christophe Nicolle[2], Yvon Voisin[2]*

[1] i3mainz, Institute for Spatial Information and Surveying Technology University of Applied Sciences Lucy-Hillebrand-Str. 2 55128 Mainz, Germany

e-mail: { boochs, andreas.marbs, helmi.benhmida, hung.truong, ashish.karmacharya }@geoinform.fh-mainz.de

[2] Laboratoire Le2i, UFR Sciences et Techniques Université de Bourgogne B.P. 47870, 21078 Dijon Cedex, France

e-mail: {christophe.cruz, Adlane.Habed, yvon.voisin, cnicolle}@u-bourgogne.fr

## ABSTRACT

Object reconstruction is an important task in many fields of application as it allows to generate digital representations of our physical world used as base for analysis, planning, construction, visualization or other aims. A reconstruction itself normally is based on reliable data (images, 3D point clouds for example) expressing the object in his complete extent. This data then has to be compiled and analyzed in order to extract all necessary geometrical elements, which represent the object and form a digital copy of it. Traditional strategies are largely based on manual interaction and interpretation, because with increasing complexity of objects human understanding is inevitable to achieve acceptable and reliable results. But human interaction is time consuming and expensive, why many researches has already been invested to use algorithmic support, what allows to speed up the process and to reduce manual work load.

Presently most of such supporting algorithms are data-driven and concentrate on specific features of the objects, being accessible to numerical models. By means of these models, which normally will represent geometrical (flatness, roughness, for example) or physical features (color, texture), the data is classified and analyzed. This is successful for objects with low complexity, but gets to its limits with increasing complexness of objects. Then purely numerical strategies are not able to sufficiently model the reality.

Therefore, the intention of our approach is to take human cognitive strategy as an example, and to simulate extraction processes based on available human defined knowledge for the objects of interest. Such processes will introduce a semantic structure for the objects and guide the algorithms used to detect and recognize objects, which will yield a higher effectiveness. Hence, our research proposes an approach using knowledge to guide the algorithms in 3D point cloud and image processing.

## 1. INTRODUCTION

As object reconstruction is an important task for many applications considerable effort has already been invested to reduce the impact of time consuming manual activities and to substitute them by numerical algorithms. Most of such algorithmic conceptions are data-driven and concentrate on specific features of the objects, being accessible to numerical models. By means of these models, which normally describe the behavior of geometrical (flatness, roughness, for example) or physical features (color, texture), the data is classified and analyzed.

One common characteristic of such strategies is to be static and not to allow dynamic adjustment to the object or to initial processing results. An algorithm will be applied to the data and producing better or minor results depending on several parameters like image or point cloud quality, completeness of object representation, position of view points, complexity of object features, use of control parameter and so on. Mostly there is no feedback to the algorithmic part in order to choose a different algorithm or just the same algorithm with changed parameters. This interaction is mainly up to the user who has to decide by himself, which algorithms to apply for which kind of objects and data sets. Often good results can only be achieved by iterative processing controlled by a human interaction.

With increasing complexity of the data and the objects represented therein, a correct validation of numerically modeled features gets again more difficult, why decisions based on individual algorithmic features tend to be unreliable. This problem only can be solved when further supplementary and guiding information is integrated into the algorithmic process chain allowing to support the process of validation. Such information might be derived from the context of the object itself and its behavior with respect to the data and/or other objects or from a systematic characterization of the parameterization and effectiveness of the algorithms to be used. For the processing such information gets accessible by rules, which will be integrated into the procedure and thus put semantic characteristics into the processing chain. Conventional programming languages support such semantic by logical or numerical conditions.

But as programming languages used in the context of numerical treatments are not dedicated to process knowledge their use of conditions is inflexible and makes the integration of semantic aspects difficult. Perhaps that's why up to now a combined processing of knowledge and numerical aspects is not very common and cannot be found in practical solutions.

But this situation may change with new technologies coming up in the framework of the semantic web. One of those technologies is a language that helps to define ontologies; an evolved version of the semantic networks. Ontologies represent one of the most famous technology for knowledge modeling, where the basic ideas was to present information using graphs and logical structure to make computers able to understand and process it easily and automatically [1].

With the work presented here, we try to build a bridge between such semantic modeling and numerical processing strategies. This avoids actual limits in the use of knowledge within numerical strategies. As base for such an approach available knowledge will be structured and explicitly formulated by linking geometrical objects to semantic information, creating rules and finally guiding the algorithms used to process the real data. The created knowledge will be structured in ontologies containing a variety of elements like already existing information to the objects as can be taken from other data sources (digital maps, geographical information systems,...) or information about the objects characteristics, the hierarchy of the sub elements, the geometrical topology, the characteristics of processing algorithms etc.

During processing, such modeled knowledge provides relevant information allowing to guide the analysis and the identification process. This will even allow to freely choosing between different algorithmic strategies, to combine them and to react onto unexpected situations by making use of the overall knowledge framework. To achieve this, all relevant information about the objects, the algorithms and their interrelation has to be modeled inside the ontology, including characteristics like positions, geometrics information, images textures, behavior and parameter of suitable algorithms, for example.

The following paper is structured into section 2 which gives a overview to actual existing strategies for reconstruction processes, section 3 explains the framework of a knowledge based approach, section 4 shows different strategies and level of knowledge for the processing, section 5 gives first results for a real example and section 6 concludes and shows next steps planned.

## 2. RECONSTRUCTION STRATEGIES

### 2.1. Manually supported strategies

In practice, tools used for 3D reconstruction of objects are still largely relying on human interaction. The largest impact of manual activities can be found in those instruments and software packages designed as construction tools with integrated viewer for different types of data sets, like images or point clouds. Here the user might be supported in his construction activity, but object interpretation, selection and extraction of measurements has to be done completely by the user. That's why this processing is the most time consuming way to come from a data set to extracted objects.

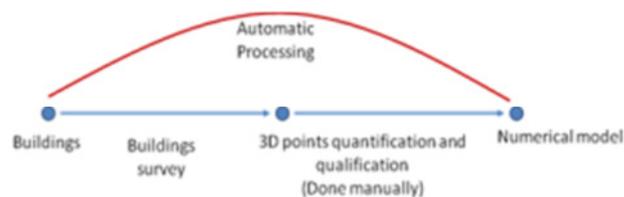

**Figure 1.** Automatic processing compared to the manual one

A first optimization is provided by semi-automatic methods. Here, the user initializes the process by some manual measurements based on which an algorithm tries to extract other elements. Algorithms support certain geometrical processing or classifications initiated by the human operator. Especially for buildings or other regularly shaped objects like man-made facilities several numerical simplifications and conditions are possible. For example, buildings can be regarded as compositions of a few components with simple roof shapes (such as flat roofs, gable roofs and hip roofs). Vosselman et al. [2], [3] tried to reconstruct a scene based on the detection of planar roof faces in the generated point clouds based on the 3D Hough transform. The used strategy relies on the detection of intersection lines and height jump edges between planar faces. Once done, the component composition is made manually.

### 2.2. Automatic strategies

Automatic methods process the data without the need of any kind of user intervention and use mainly various segmentation techniques to extract features. Pollefeys et al. [3] and Hartley et al. [4] show examples for strategies based on projective geometry. [3] combines various algorithms from computer vision, like projective reconstruction, auto-calibration and depth map estimation. The disparity calculation between point pairs makes it possible to get a depth map. The depth map is then transformed into a volume model composed of voxels. The surface estimation between the outer surface voxels and the interior surface voxels makes it possible to combine inner and outer object parts. The method developed is effective and obtains good results. The approach of Hartley et al. [4] proceeds in two steps. First, a coarse surface model of the building is carried out. Then the coarse model guides the search of details (windows and doors) and refines the surface model. The reconstruction uses the detection of "vanishing points", line correspondence, and the estimation of points and homologous lines. Vanishing points are necessary for the detection of planar primitives with the help of the plane-sweeping method. This method has strong constraints as it contains three perpendicular dominant directions.

Although these methods are already successful, the degree of incompleteness or erroneous decisions is too high for an integration into practical work. On the other hand, in case of practicability they would considerably reduce manual work load.

**2.3. Strategies with knowledge support**

Improvements for automatic processing can be expected from new strategies based on semantic networks used to guide the reconstruction like the work of Cantzler et al. [5] or Scholze et al. [6]. They use certain architectural features like orientations of a wall, for example. The whole strategy consists of three steps. First architectural features are extracted from a triangulated 3D model, then constraints are generated out of the scene by matching planes against a semantic of the building mock up by a backtracking research tree. In this step, the semantic network concentrates on the definition of the 3D objects and the relationships among them. Constraints such as parallel or perpendicular to a wall are exploited. Finally constraints found are applied, what then allows to extend and update the original model. Scholze et al. [6], extend this work into a model based reconstruction of complex polyhedral building roofs. The roofs in question are modeled as a structured collection of planar polygonal faces. The modeling is done into two different layer, one focus on geometry, whereas the other is rules by semantics. Concerning the geometry layer, the 3D line segments are grouped into planes and furthers into faces using a Bayesian analysis. In the second layer, the preliminary geometric model is subject to a semantic interpretation. The knowledge gained in this step is used to infer missing parts of the roof model (by invoking the geometric layer once more) and to adjust the overall roof topology. This work exemplarily shows the potential of semantic rules taking relations between certain characteristics into account. But although the rules used here were mainly simple, semantic tools meanwhile offer a broad framework to combine geometrical, topological, factual or logical aspects.

**2.4. Elements of a new strategy**

The problem of automatic object reconstruction remains a difficult task to realize in spite of many years of research [7]. Major problems result from geometry and appearance of objects and their complexity and impact on the data collected. For example, variations in a viewpoint may destroy the adjacency relations inside the data, especially when the object surface shows considerable geometrical variations. This dissimilarity affects geometrical or topological relations inside the data and even gets worse, when partial occlusions result in a disappearance of object parts. Efficient strategies therefore have to be very flexible and in principle need to model almost all factors having impact of the representation of an object in a data set. That leads to the finding, that at first a semantic model of a scene and the objects existing therein is required. Such a semantic description should be as close to the reality as possible and as necessary to take most relevant factors into account, which may have impact on later analysis steps. At least this comprises the objects to be extracted with their most characteristic features (geometry, shape, texture, orientation,...) and relations among each other. The decision upon features to be modeled should be affected by other important factors in an analysis step like characteristics of the data, the algorithms present and their important features.

Such a model might be expressed by a semantic network formed out of nodes and connections. The nodes represent classes or objects as their instances and the links show relationships of various characteristics. Such a network then contains the knowledge of that type of scene, which has to be processed. This knowledge base will act as basis for further extraction activities and has to work in cooperation with numerical algorithms.

Up to this point, the new conception is still in concordance to other knowledge related set ups, although the degree of modeling goes farther because all relevant scene knowledge will be integrated. But another aspect will be considered also allowing to considerably improve processing strength. That is to integrate knowledge even on the algorithmic side. This means to make use of the flexibility of knowledge processing for decisions and control purposes inside the algorithmic processing chain. Even a propagation of findings from processing results into new knowledge for subsequent steps should be possible, what would give a completely new degree of dynamics and stability into the evaluation process.

Consequently a further knowledge base has to be developed which characterizes algorithms, their relation among each other and their relation to the scene knowledge. As a result, the processing will no more be guided by the numerical treatments and their results but by the a complete knowledge base comprising all available semantics, including scene knowledge, object knowledge, algorithmic knowledge and important relations among them. This then leads to a conceptual view as shown in figure 2. The processing named as WiDOP (knowledge based detection of objects in point clouds and images - in correspondence to a project of the same name) has its roots in the knowledge base which then guides individual algorithmic steps. The decision upon these algorithmic steps is taken from the knowledge base. Results from algorithms are also analyzed by the knowledge base and the reasoning engine, then deciding upon subsequent steps. Accordingly detected objects and their features are populated to the knowledge base, which will permanently increase until the work is done.

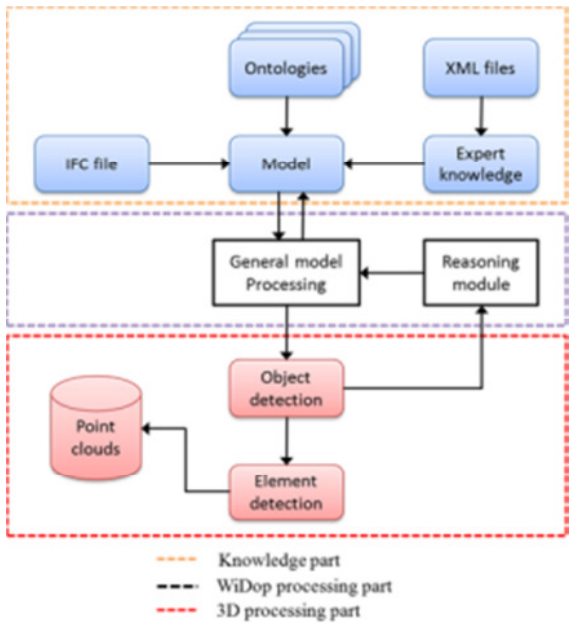

**Figure 2.** WiDOP system overview

As base for such a strategy developments for the semantic Web and semantic technologies inside will be used. There are languages available allowing to define ontologies, which contain available knowledge to a certain scenario [1]. This knowledge then has to be structured and prepared to guide the algorithmic processing. As input sources general expert knowledge may serve, but also other resources like GIS data, information to object characteristics, to hierarchical structure of scenes, to the geometrical topology, or to different processing algorithms etc. Inside the automatic detection process, the modeled knowledge will provide all relevant information necessary to guide the localization and the identification process. How this can be achieved will be explained in the following sections.

## 3. INTEGRATION OF KNOWLEDGE INTO A DETECTION STRATEGY

### 3.1. Overall strategy

Following to above considerations and with respect to technological possibilities, knowledge will be modeled in various levels. In principle we have to distinguish between object-related knowledge and algorithmic knowledge and we therefore have a

- Layer of object knowledge
- Layer of algorithmic knowledge

containing the respective semantic information already explained before.
Object knowledge will be classified in three categories: geometric, topologic and semantic knowledge representing a certain scenario [8] Therefore we distinguish between:

- Scene knowledge
- Geometric knowledge
- Topological knowledge

The layer of scene knowledge contains all relevant object elements which might be found within that scene. In case of buildings, this might comprise a list like: {Building, Wall, Door, Window, Ground,…}.
Geometrical knowledge formulates geometrical characteristics to the physical properties of scene elements. In the simplest case, this information might be limited to few coordinates expressing a bounding box containing the element. But for elements being accessible to functional descriptions this will extend the description. A wall, for example, has a vertical plane, which needs to be described by a plane equation, its values and completed by width and height.
Topological knowledge represents adjacency relationships between scene elements. In case of a building, for example, a topological relation between a wall and the ground floor can be defined, as both have to be connected and the wall must be perpendicular to the ground.
Finally the algorithmic layer contains all relevant aspects needed to guide the processing itself and expresses processing sequences, data exchange necessary, parameterization of individual algorithms and relations to other layers. The integration of 3D processing algorithms into the semantic framework is done by means of special Built-Ins called "Processing Built-Ins". They manage the interaction between above mentioned layers and will be explained later on.

### 3.2. Use and modeling of knowledge

The framework to express and access knowledge by a computer is provided by ontologies and tools to handle them in a software environment. An ontology is a formal representation of knowledge as a set of concepts within a domain, and the relationships between those concepts. It is used to reason about the entities within that domain, and may be used to describe the domain. In theory, conventionally, ontology presents a "formal, explicit specification of a shared conceptualization"[8]. An ontology provides a shared vocabulary, which can be used to model a domain. Well-made ontologies own a number of positive aspects like the ability to define a precise vocabulary of terms, the ability to inherit and extends exiting ones, the ability to declare relationships between defined concepts and finally the ability to infer new relationships by reasoning on existing ones.
In the context of 3D processing, formal ontologies have already been suggested as a possible solution to reconstruct objects from 3D point clouds [5], [6], [7]. Through technologies known as Semantic Web, most precisely the Ontology Web Language (OWL) [9], researcher are able to share and extends knowledge through the scientific community. Lots of reasoners exist nowadays like Pellet [10], Fact++ [11], and KAON[12]. They use rules to perform particular operations on knowledge bases like the consistency checking, the satisfiability checking and finally the expansion of relationships between objects inferred from explicitly stated relationships.

Despite the richness of OWL's set of relational properties, it does not cover the full range of expressive possibilities for object relationships that we will need, since it is useful to declare relationship in term of conditions or even rules. These rules are integrated through different rule languages to enhance the knowledge existing in an ontology. In the last few years, lots of rule languages have been emerged. Some of the evolved languages are related to the semantic web rule language [13] (SWRL) and advanced Jena rules [14]. SWRL is a proposal for a Semantic Web rules-language, combining sublanguages of the OWL Web Ontology Language (OWL DL and Lite) with those of the Rule Markup Language [13]. In addition, these languages are open and flexible and allow to integrate Built-Ins, which in our case give access to the world of geometrical processing.

A simple example rule would be to assert that the combination of the *hasParent* and *hasBrother* properties implies the *hasUncle* one. This rule could be written as:

*hasParent*(?x1,?x2) ^ *hasBrother*(?x2,?x3) → *hasUncle*(?x1,?x3)

Where x1, x2 and x3 present individuals from the class *Person* defined in the ontology and *hasParent*, *hasBrother* and *hasUncle* presents data properties in the same cited structure. As seen in the above example, rules are divided in two parts, antecedent and consequent separated by the symbol "→". If all the statements in the antecedent clause are determined to be true, then all the statements in the consequent clause are applied. In this way, new properties like *hasUncle* in our example can be assigned to individuals in the ontology based upon on the current state of knowledge base. Add to this standard, SWRL language specify also a library for Built-Ins functions which can be applied to individuals. It includes numerical comparison, simple arithmetic and string manipulation.

### 3.3. Knowledge related to the object
#### 3.3.1. Scene layer classes
Figure 3 shows a possible collection of scene elements in case of a building. They may be additionally structured in a hierarchical order as might be seen convenient for a scene. This could lead to relations like: a room is a super class of wall and floor, with door as further sub class of wall. But also other ordering can be imagined, as a structuring with respect to processing aspects. Such a structure could separate between different complexity of elements. Simply structured elements like walls, ground floors or ceiling then would be distinguished from other objects in accordance to their impact onto the processing strategy. Simple objects for example will inherit simple geometries (like planes, for example) and correspondingly only need simple detection strategies, whereas complex ones will be composed out of several geometrical elements needing adapted and more complex processing strategies. They first have to be decomposed into their geometric elements, which then have to be verified and regrouped based on known topology relations between them. Likewise, a table as a complex element is composed of a plane representing the table top and at least one linear structure, representing a leg. Once theses geometries are detected, the topology decides upon the correctness of this assumption, as the plane (table top) must be connected and perpendicular to the linear structure (leg). This is just a first draft since modeling depends on the target scene to be detected.

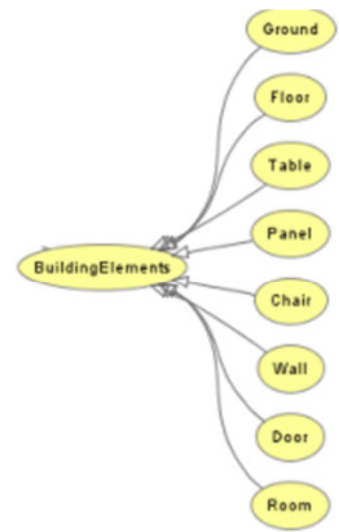

**Figure 3.** Grouping of scene elements in case of a building

Above cited concepts are extended by relations to other classes or data. The data property "*has_Bounding_Box*" for example aims to store the placement of the detected object in a bounding box defined by its 8 corner points (a spatial point is defined by 3 values x, y and z). The object property "*has_Geometric_Component*" aims to specify the geometric elements composing the semantic object in question.

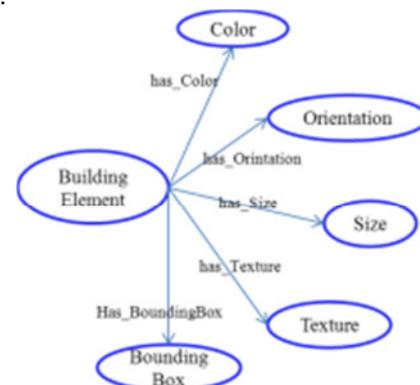

**Figure 4.** Object and data properties characterizing the semantic objects

To specify its semantic characteristics, a new class are created, aiming to characterize a semantic object by a set of characteristics like color, size, visibility, texture, orientation and its position in the point cloud after detection. To do so, new object properties like "*has_Color*", "*has_Size*", "*has_Orientation*", "*has_Visibility*" and "*has_Texture*" are created linking the *Semantic_Object* class to the "*color*", "*size*", "*Orientation*", "*Visibility*" and "*Texture*" classes respectively, cf. Figure 4.

*3.3.2. Geometry layer classes*
Each one of the above mentioned object classes have relations to the geometry class. This class handles features which may have an impact or are useful for decisions based on geometrical aspects. It helps to enrich scene objects with additional information or provide data for the processing strategies. A basic couple of sub-classes labeled "*Geometric_Component*" and "*BoundingBox*" are created, cf. Figure 5.

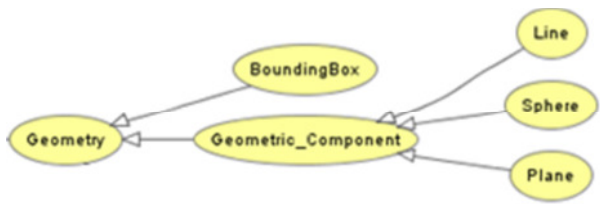

**Figure 5**. The geometry class hierarchy

- *Geometric_component* class:
  The *Geometric_Component* class contains information about the different geometric elements composing a semantic object, like plane, line, sphere and others. A wall for example has a planar geometry; moreover, a table consists of planar and linear geometries.
- *BoundingBox* class:
  The BoundingBox class aims to characterize the object localization within the 3D point cloud scene.

*3.3.3. Topology layer classes*
The purpose of this class is to spatially connect *Things* presented in the scene and in the geometry layer class. At semantic view, topological properties describe adjacency relations between classes. For example, the property *isParallelTo* allows to characterize two geometric concepts by the feature of parallelism. Similarly relations like *isPerpendicularTo* and *isConnectedTo* will help to characterize and exploit certain spatial relations and make them accessible to reasoning steps.

**3.4. Processing layer classes**

All above mentioned layers carry knowledge, which describes the objects in their semantic and spatial context. This knowledge is the source to reason and decide upon findings and results produced by individual 3D-processing steps. The intrinsic 3D-processing, however, is done on a different level, due to several reasons.
One of these reasons is the need to efficiently and quickly process large data sets (images, point clouds) what must be realized with accordingly designed programming tools. That's why above mentioned framework for knowledge management cannot be used. Its structure is not designed for that task. Instead, an interface has to be implemented giving access to efficient programming languages, like C or C++, for example. Fortunately, Java provides all necessary structures to build such an interface and therefore acts as bridge to combine these real different worlds of semantic processing and efficient data processing.

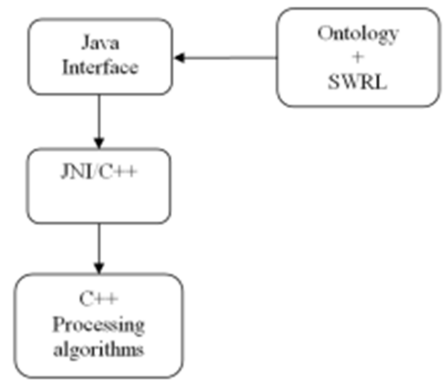

**Figure 6.** Processing architecture

As shown in the figure 6 above, JAVA has interfaces to the semantic world and also to the processing world. As consequence, it is possible to start and control activities inside the processing environment based on functions implemented in the semantic framework.
In the simplest case, this would allow to define a C or C++ function representing a defined sequence of processing steps and to start this function as a semantic built-in method. In that case the 3D-processing wouldn't do more than just return information to be used in a further reasoning process on the knowledge level. However, this is already useful to exploit the potential of knowledge management for the guidance of 3D-processing.
However, such a solution of defining individual processing sequences and connecting them to an own spatial built-in method is of limited flexibility. It would need a large number of methods representing a complete tool box covering most of the possible processing situations. This might result in a certain redundancy between processing built-ins for similar objects or for the same object to be analyzed under different conditions.
A higher degree of flexibility and less redundancy could be achieved by developing an own processing semantic and to exploit this by the reasoning capacity inside the knowledge processing. Such a solution would need to describe each individual algorithm by features, which are characteristic and important to model its behavior. This information then could be treated as key to reason about usefulness of a certain algorithm for a specific detection situation. The reasoning would have to be based on features of the objects in the scene and their importance for a processing decision. Thus, the semantic inside the processing network has to be attached by relations and rules to the scene knowledge.

Such an extended and more flexible connection between scene and processing domain needs extended experience with algorithms and their interaction with certain characteristics of objects and data. That's why an implementation has to wait until the experience needed could be collected based on a realization of the built-in solution, explained at the beginning.

## 4. THE IMPACT OF KNOWLEDGE ON THE DETECTION STRATEGY

### 4.1. Use of knowledge in general

The previous chapter gave an idea to detection strategies in general, to a possible integration of knowledge into a detection process and to concepts of structuring knowledge in order to support algorithmic processing. The next question to answer is how the overall detection strategy might be influenced by knowledge and what this means for the design of a practical solution.

As explained before, knowledge is the key element in this solution and it has to guide and control the process of detection. It has to be stored and organized in a specific way, in order to be accessible for the reasoning process. Our proposition has been shown in chapter 3, but other structures are also possible. One aspect not considered up to now is how knowledge may guide the processing and to what extend it might be necessary to distinguish different degrees of available knowledge. This will be done in the following chapter, explaining two major strategies:

- use of well defined specific knowledge
- use of generic knowledge

Why this difference? We have to accept, that each individual application case has its own framework of knowledge. The content of such a framework changes with the domain to which an application has to be referenced (architecture, industry, civil engineering,....) and accordingly knowledge models to be used must be different. In addition, the framework will be influenced by the amount of knowledge existing in a particular application. This may spread a large field, starting from extensive and actual data bases with more or less precise information up to just some general ideas to objects in question and without any direct data on the other end.

Such large differences in the knowledge base clearly must have impact on the guidance of algorithms and on the strategies used. In principle, the more knowledge existing, the more precisely and directly algorithms can be guided, why there are strategically different concepts following the degree of quality for the knowledge. Hence we distinguish between sparse knowledge cases (generic knowledge, cf. Figure 7, left side ) and detailed knowledge cases (specific knowledge, cf. Figure 7, right side) .

### 4.2. Case of specific knowledge integration

#### 4.2.1. The role of specific knowledge

What are potential data sources for 'specific knowledge' and why might it still be necessary to analyze data sets and look for objects?

Data sources are various. This might range from simple CAD plans over spatial information systems to object oriented data bases supporting data in rich and complex formats like IFC [18]. Based on these data sources the different levels (scene, geometry, topology) in our knowledge model can be expressed as far as possible. In an ideal case we therefore might know about the semantic of objects (there are walls, floors, ceiling,...), the geometry (position, extension, orientation,...), additional features (roughness, color, other surface characteristics) and topological relations (wall A sits on floor B), what would give a really good base for a detection strategy. This knowledge then has to be linked to the algorithmic knowledge, what finally would allow to start and guide the processing part (see 4.2.2.).

As soon as the degree of completeness and particularity of the data sources diminishes, the knowledge gets more and more sparse and needs to adapt the strategy. One first change arises with lack of knowledge to the geometry of objects (see 4.3.).

But what are reasons to detect objects which are already known to a more or less detailed degree? It's necessary due the fact, that data sets have a certain age and actuality, which in many cases don't match the needs to be fulfilled by the data sets.

Official topographical data sets have an age between 1 and 20 years and even continuously evolving objects like industrial plants may have data sets of similar age. But even for younger data sets an analysis could be of interest as contained objects may undergo permanent changes.

One example for such a type of objects is an airport. Everybody should already have noticed all the construction sites being permanently visible on the airports all around the world. Building parts, elements of infrastructure are undergoing many changes. Walls are disappearing, new walls are showing up, new openings or closings inside walls arise or elements of various technical infrastructures get modified. Normally those changes are not updated into the data bases, why they suffer an increasing lack of quality and actuality. An airport might seem as a special example, but there are many similar scenarios for aged data sets.

One reason for missing updates of data bases are the costs arising. In practice, update measurements are done manually and due the amount of time to be invested it gets very expensive. That's why an automatic solution just looking for existence or disappearance of objects noted in a data base already could give a large economical progress.

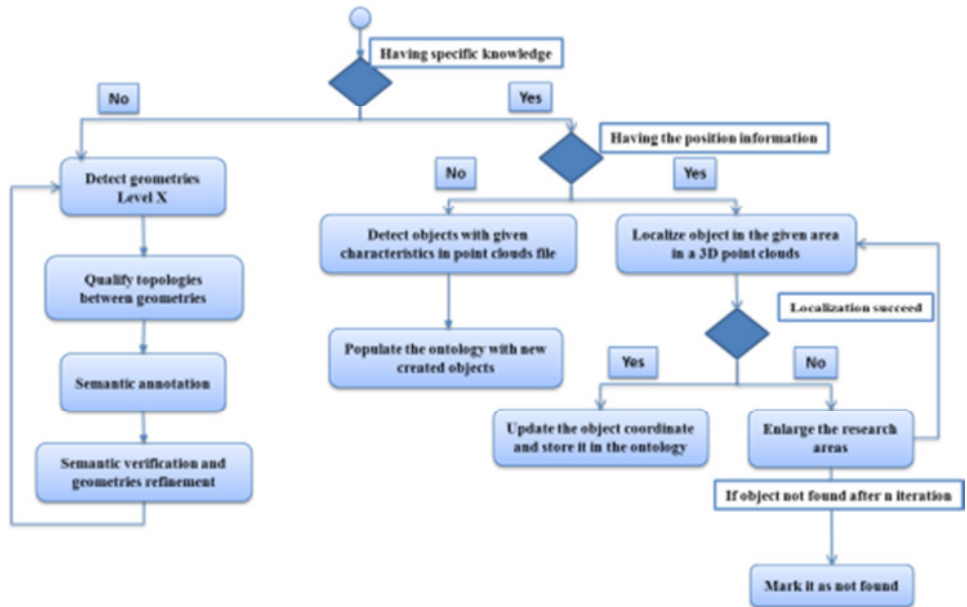

**Figure 7.** WiDOP activity diagram

*4.2.2. Processing in case of detailed knowledge including object position*

This case represents the ideal situation from the view point of existing knowledge. Remaining challenges for the guidance of the processing come mainly from the data to be analyzed, possible incompleteness, lack of data quality, for example and the algorithmic knowledge needed to handle such situations.

Figure 8 presents the adopted strategy in this case using point clouds as data source.

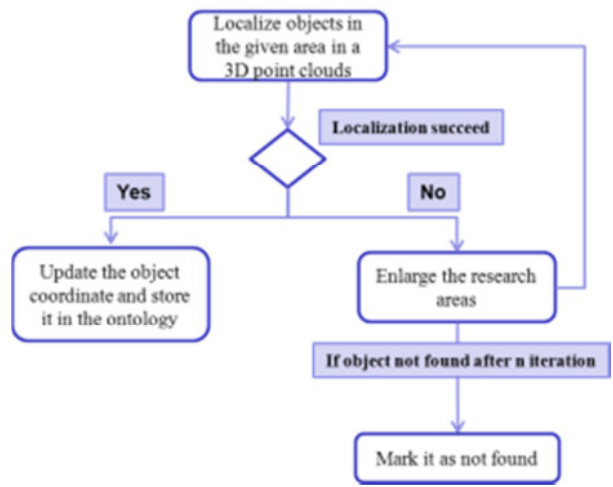

**Figure 8.** Activity diagram in case of specific knowledge with known object position

The first step localizes the target object in the data set (for example, a point cloud) based on previously mentioned 3D_Processing_Built-Ins (see **3.4**). It aims to infer knowledge and execute one or more 3D processing algorithms with extracted knowledge from the ontology. Once the localization is done successfully, the object will be stored within its coordinates in the ontology. In case of a failure, the algorithmic knowledge has to decide upon the next step, what could be an enlargement of the research area. Such a step would assume, that the reason for the failure is due to imprecise geometry data, why the process of localization should be re-executed. Finally, the object coordinate can be updated in case of a successful localization, if not, it will be marked as not found, or further rules have to be applied.

**4.3. Case of generic knowledge integration**

With decreasing knowledge, especially with lack of prior information to a position, the processing has to use a largely different strategy. In previous cases of an available position, this guides a detection process spatially and semantically. This is because the position guides the process to a specific location in the data set; and as the used geometry is bound to a certain object, the semantic part is already decided. The remaining task of the algorithms is just to decide upon correctness of the assumptions provided by the knowledge.

But without geometry a direct link between an object and its corresponding representation in the data does not exist. That's why geometry and semantic information are undetermined at the beginning. Subsequent algorithmic steps then have first to localize something in order to solve the geometry question and afterwards to decide upon the semantic. These decisions are interdependent and have to be taken in a concerted way based on generic knowledge. The quality of this knowledge is decisive for the effort to invest and the quality and remaining uncertainness in the detection process.

Looking from a procedural perspective, fig. 9 shows a corresponding strategy. Here, each iteration is composed of four different steps. The first tries to detect basic geometrical elements, which may be part of a physical object (like planes, lines, for example). At that moment geometry information is available, but it is unclear to which object the elements found may belong. This has to be answered using a different generic logic, as may be derived from topology, for example. Thus a next step verifies topological relations between detected elements and adds other aspects like orientation (vertical element, horizontal element,...). Based on results from this reasoning a semantic annotation process can be executed in order to obtain an initial mapping between elements derived from the data and the generic semantic.

Such a mapping extends the knowledge in the ontology from generic to specific one as now real objects have been created. However, this specific knowledge might be of lower quality or of higher uncertainty than in case of precise initial information from external sources. Therefore a further step has to be added, allowing to improve the quality. One way to achieve an improvement is to use the processing chain for specific knowledge and to apply those algorithms, which need a closer object context but give better results.

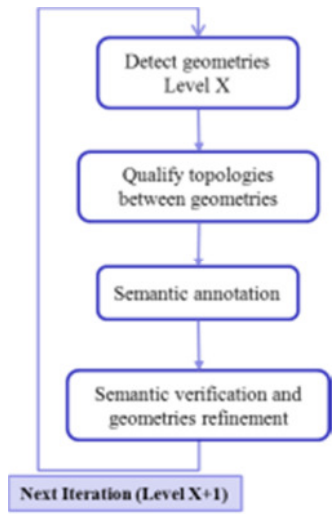

**Figure 9.** Activity diagram in case of generic knowledge

A successful detection then may lead to a subsequent refinement process, allowing to identify less prominent objects, which are smaller or more complex and therefore need more support for an identification. This may even be simply based on generic knowledge, providing general concepts to objects and their relation among each other. As example, it is clear, that a table has to sit on a ground floor and that chairs may have close adjacency to other chairs or to tables. As consequence, generic knowledge may guide the detection process in an iterative way, leading from large and significant objects to smaller and more complex ones.

## 5. CASE STUDY

In the following chapter we want to describe and explain the implementation of our concept based on real data of an airport scene (Frankfurt Airport = Fraport), we are using in our ongoing development procedure at the moment.

On the knowledge side, an ontology is created, using the open-source ontology editor protégé with Web Ontology Language (OWL) [9]. Rules will be formulated and executed using the Semantic Web Rule Language (SWRL) [13] and its built-ins, where the reasoning is performed with SWRL Jess tab and Jena. In addition, new 3D processing Built-Ins are developed dealing with real 3D-point cloud data and image processing tasks. These algorithms are separated functions, programmed in C++. They are modularly structured allowing to use them individually or to be grouped together, depending on the purpose of use. The semantic part plays his role as control logic for the algorithms, setting parameters as well as evaluating the results in order to make further decisions (cf.).

In order to go into detail of this approach, an example of the defined Built-Ins will be explained in the following paragraphs, then, a rule based model related to the target purpose will be mentioned for each type of existent knowledge, either specific or generic.

### 5.1. 3D scene definition

The Fraport scene, which is used as a first exemplary dataset for the implementation, is an indoor architecture of a waiting room in a boarding area of Frankfurt airport. It contains regular walls, floor, chairs, advertisement panels, signs etc. The whole scene has been scanned using a terrestrial laser scanner, resulting in a large point cloud representing the surfaces of the scene objects captured from different scanning positions. Based on this data and the knowledge defined we will try to give a clear understanding of benefits from a knowledge based 3D processing, in the following.

### 5.2. 3D processing Built-Ins: plane detection

As explained previously (cf.3.4), so-called processing Built-Ins provide the bridge between semantic and processing world. They are defined as part of semantic rules and have a counterpart in the processing domain, which has to be invoked as the rule is executed. 3Dswrlb:PlaneDetection Built-In is one example for such a processing procedure. It aims to detect planes within 3D-point clouds based on certain characteristics of the target object properties. In this context, we will define the *PlaneDetection* Built-Ins as seen in Figure 10. The prototype of the designed Built-Ins is:

```
3D_swrlb_Processing:Plane_Detection(
Building    Element,    Orientation,
Texture, Thickness, Height, Position,
Box)
```

The first parameter represents the target object if available, and the last one represents the created bounding box once this geometry element is detected. The remaining parameters are target object characteristics, used as input information for the 3D processing, in order to choose the correct algorithmic strategy and parameterization. At the moment, each object detection process will result in a bounding box, representing a rough position and orientation of the detected object for visualization purposes. For the future a parametric object description will be used, for example using just a point and a normal vector for the geometric object "plane" or start point, end point and height for the semantic object "wall".

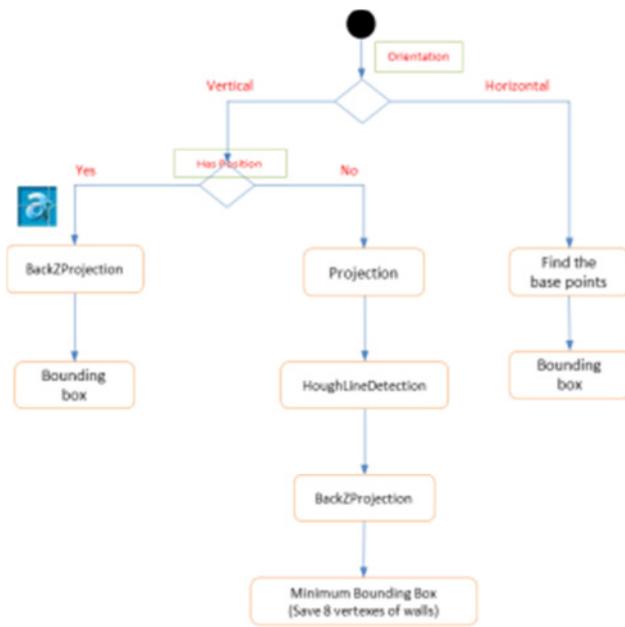

**Figure 10.** Plane detection Built-Ins execution

In figure 10, the orange boxes represent individual 3D-processing functions, used for the plane detection. Each branch then represents a different processing chain, applied in case of horizontal or vertical planes with or without known position. Each processing chain uses functions appropriate to the knowledge available. In the left hand branch, for example, the object position is expected to be known with a certain accuracy why function "BackZProjection" may use known 2D coordinates to pre-segment the 3D points within the point cloud and finally create a bounding box. In that step it is evaluated if the segmented points belong to a plane or not.

In the central branch we know the target object ("wall") but not the position, why the whole point cloud will first be projected to the ground view plane in order to get a dense contour at the position of walls(function "projection"), then a Hough line detection algorithm will detect possible vertical elements with a certain length before the "BackZProjection" function will again segment 3D points. Finally, based on processing results and scene knowledge, it will be decided, which candidates belong to a wall and the bounding boxes will be created as a processing result.

In the right hand branch, the ground floor of the waiting area will be detected. Here an algorithm is used that will shift a thin horizontal search space from bottom to top, stopping when a maximum number of 3D points is with the search space. The result will be described by a bounding box.

### 5.3. Case of specific knowledge integration

Above mentioned algorithmic processing has semantic counterparts integrated into rules. That's why there has to be a semantic framework choosing one of the processing alternatives. For example, in order to detect walls with known position (left hand branch in figure 10) we need certain rules that may look like rule demo1 below. For this it is necessary, that all objects with specific knowledge already exist as individuals in the ontology, containing all available information like position, color, size or orientation.

In plain words, rule1 could read like this: "Select all individuals of the class 'wall' in the ontology which have a known position and a height larger than 4 meters and execute the adapted plane detection algorithm for this case. If successful, store the resulting bounding boxes for each wall in the ontology."

Rule1:
```
Wall(?x) ^ has_Position(?x,?pos)
^ 3D_swrlb_Processing: Plane_Detection
(?x, Orientation:Vertical,
texture:Flat, Thickness:Thick, Height:
>4m, Position:?pos , ?box) ^
hasDetectionRes (Wall,True)
 → hasBoundingBox(?x,?box) ^
hasQualification (?box,Semantic) ^
hasPosition(?box, ?pos)
```

Figure 11 shows the result of the wall detection, using rule1.

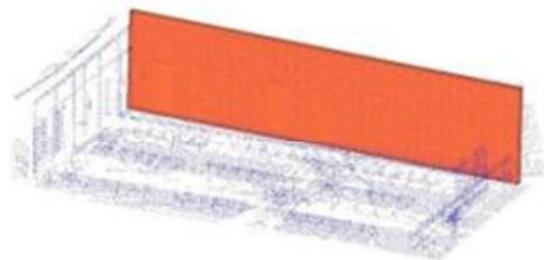

**Figure 11.** Related result to rule1

To satisfy the case where no specific information related to the target object (wall in our case) is given, rule2 is designed:

Rule2:
```
swrlb_Processing: Plane_Detection
(?Wall, Orientation: Vertical,
texture:Flat, Thickness: Thick,
Height: >4m, Position:any, ?box) →
hasBoundingBox(?Wall,?box) ^
hasQualification (?box,Semantic) ^
hasPosition(?Box, xxx)
```

With this rule, all vertical planes with a height larger than 4 meters will be detected, using an adapted plane detection strategy (central branch in figure 10). After the execution of the plane detection Built-Ins with the given attributes, the processing will result in a certain number of bounding boxes, representing the walls. Once done, the rule will be converted to rule3:

Rule3:
```
BoundingBox (?Box) ^ hasDetectionRes
(?Box, True) → Wall(?box) ^
hasQualification (?Box, Semantic) ^
hasPosition(?Box, ?pos)
```

## 5.4. Case of generic knowledge integration

In this section we want to show, how knowledge rules are designed for reasoning if no specific knowledge is available. To do so, different rule steps are necessary.
First, all vertical planes of certain characteristics will be searched in the area of interest, then topological relations between elements found are analyzed and are used to further qualify an element. Subsequently further rules may be applied containing generic aspects expressing facts to orientation or size of elements, which then may already be sufficient to finally decide upon the semantic of an object.

*5.4.1. Detect geometry*
```
swrlb_Processing:Plane_Detection
(?Geometry, Orientation:Vertical,
texture:Flat, Thickness:any, Height:
<4m, Position:any, ?box) →
hasQualification(?box, "geometric")
^hasPosition (?box, xxx)
```

With this rule all vertical building elements where the height is less than 4 meters, are detected and stored in the ontology as a geometric element.

*5.4.2. From geometric to topologic level*
This step aims to identify existing topologies between the detected geometries. To do so, 3D_Processing Built-Ins like *Perpendicular* and *Connection* are created. Each Built-Ins accesses certain 3D functions, programmed in C++, which will verify the respective topological relations. As a result, relations found between geometric elements are propagated into the ontology, serving as an improved knowledge base for further processing and decision steps.
Rule4:
```
BoundingBox(?box1) ^
BoundingBox(?box2) ^
swrlb_Processing:Perpendicular (?box1,
? box2) → isPerpendicularto(?box1, ?
box2)
```

Rule5:
```
BoundingBox(?box1) ^
BoundingBox(?box2) ^
swrlb_Processing:Connection(?box1, ?
box2) → isConnectedto(?box1, ? box2)
```

*5.4.3. From geometric to semantic level*
After geometry and topology detection more rules are needed to qualify and annotate the different detected geometries based on the information in the knowledge base.

In the exemplary rule6, the inference capacity of SWRL allows the automatic annotation of a bounding box as a building object of the class "panel", based on the assumption, that panels are vertical objects with a height smaller than 4 meters.

Rule6:
```
hasOrientation (?box, Vertical) ^
hasHeight(?box, ?h) ^ swrlb:lessThen
(?h, 4) → Panel (?box)^
hasQualification (?box, Semantic)
```

*5.4.4. From topologic to semantic level*
The example of rule7 shows, how based on the existent topology (perpendicular) and in the context of Fraport, a logic annotation of the objects wall and ground can be made.

Rule7:
```
BoundingBox(?box1) ^ hasSize
(?box1,big)^ hasOrientation
(?box1,vertical)^BoundingBox(?box2)^
hasSize(?box2,big) isPerpendicularto
(?box1, ? box2) → Wall(?box1) ^
hasQualification (?Box1, Semantic) ^
Ground (?box2) ^ hasQualification
(?Box2, Semantic)
```

*5.4.5. Semantic qualification of new elements*

Rule8:
```
BoundingBox(?box1) ^ Ground(?gr)
isPerpendicularto(?box1, ? ?gr) ^
hasHeight (?box1, ?h) ^
swrlb:lessthan(2) → Gate_Counter
(?box1) ^ hasQualification (?Box1,
Semantic)
```

In this rule, the gate counter is semantically identified, using the knowledge that this object has a certain maximum size and a certain relative position and orientation (topology) to the previously detected ground floor (dark blue in figure 12).

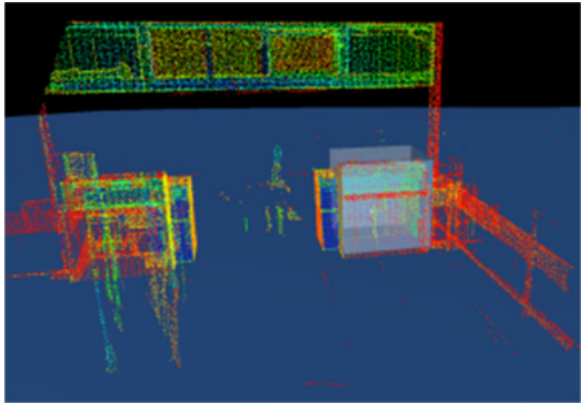

**Figure 12.** Box identified as a gate counter (light blue)

All these exemplary rules are just to show the principle of supporting 3D processing and object detection through knowledge. The rules do not present final or optimal solutions and have not been extensively tested, yet.

However, from the presented exemplary rules, we can summarize the following: Rules can be used to
- support the selection of 3D processing algorithms and their parameterization, based on scene knowledge and algorithmic knowledge
- populate the ontology based on 3D processing results in order to enrich the knowledge base to support further processing steps
- semantically annotate geometric objects based on logical reasoning
- iteratively utilize an increasing knowledge base in order to detect objects, that could not be detected initially

In a nutshell, the rules allow the minimization of the "semantic gap" between the user and the real 3D processing by means of an understandable language.

## 6. CONCLUSION AND FUTURE WORKS

This paper presents a flexible innovative solution to perform object detection in 3D data. The suggested solution makes use of available knowledge in a specific domain or scene, for example recruited from CAD drawings, and on the knowledge to respective 3D processing algorithms. This prior knowledge has to be modeled in an ontology, representing a basis for decision processed during the object detection. Semantic rules are used to control the 3D processing chain, to annotate the detected elements, to enrich the knowledge base and to drive the detection of new objects based on detected ones. The presented solution offers a flexible conception for different application scenarios, for example, for updating existing plans or reconstructing buildings based on standard "building knowledge".

Future work includes the expansion of the ontology, further implementation and testing of the rules, the improvement of the existing JAVA prototype application and the improvement and adding of 3D algorithms.

Another important aspect is the study of the quality evaluation in order to verify and improve the processing results.

## 7. ACKNOWLEDGEMENTS

This paper presents work performed in the framework of research project funded by the German ministry of research and education under contract No. 1758X09. The authors cordially thank for this funding.